# Seamless Accurate Positioning in Deep Urban Area based on Mode Switching Between DGNSS and Multipath Mitigation Positioning

Yongjun Lee, *Student Member, IEEE,* Yoola Hwang, Jae Young Ahn, Jiwon Seo, Member, *IEEE,* and Byungwoon Park, *Member, IEEE*

*Abstract*—**Multipath and non-line-of-sight (NLOS) signals are the major causes of poor accuracy of a global navigation satellite system (GNSS) in urban areas. Despite the wide usage of the GNSS in populated urban areas, it is difficult to suggest a generalized method because multipath errors are user-specific errors that cannot be eliminated by the DGNSS or a real-time kinematic technique. This paper introduces a real-time multipath estimation and mitigation technique, which considers compensation for the time offset between constellations. It also presents a mode-switching algorithm between the DGNSS and multipath mitigating mode and shows that this technique can be effectively utilized for automobiles in a deep urban environment without any help from sensors other than GNSS. The availability is improved from 64% to 100% and the error RMS is reduced from 11.1 m to 1.2 m on Teheran-ro, Seoul, Korea. Because this method does not require prior information or additional sensor implementation for high-positioning performance in deep urban areas, it is expected to gain wide usage in not only the automotive industry but also future intelligent transportation systems.**

*Index Terms*— **Deep urban area positioning, differential GNSS, global navigation satellite system, multipath, non-line-of-sight error**

## I. INTRODUCTION

Global navigation satellite system (GNSS) is the underlying technology of intelligent transportation system (ITS) [1], which is widely used in various fields such as autonomous vehicles [2] and unmanned air systems (UASs) [3][4]. As more than two-thirds of the world population will live in urban areas by 2050 [5], GNSS-based services are forecasted to be increasingly offered in urban environments [2]; therefore, the demand for improving the GNSS accuracy in urban canyons is gradually increasing. However, the positioning performance in urban canyons is inevitably much worse than that under the open sky because GNSS positioning is neither reliable nor accurate when satellite signals are blocked and/or reflected [6]. Accurate positioning in urban areas is a long-standing problem of the GNSS. When the reflected signals are received together with the direct signal, they interfere with the signals received directly from the satellites [7]; this phenomenon is generally called multipath. In addition to the typical phenomenon of multipath, non-line-of-sight (NLOS) cases in which the direct signal is blocked while only the reflected signal is received [7] frequently occur in urban areas. In this study, both cases are called multipaths for convenience.

Multipath is a site-dependent error similar to receiver noise, which cannot be eliminated by a differential technique such as real-time kinematic (RTK) or differential GNSS (DGNSS) [8]. Unlike the common GNSS errors removable through differential methods, the effect of the site-specific error is sensitively dependent on the signal reception environment. Reflective surfaces also affect the signals, which can be scattered [9] or diffracted [10] rather than simply being reflected. Thus, it is very difficult to mitigate or model the multipath, which makes it a dominant source of error in urban canyons. When GNSS signals are prone to reflection due to the presence of numerous vehicles and buildings, these effects may cause a positioning error of approximately 100 m in urban canyons [11]. The NLOS-type multipath makes receivers perceive the measurement that includes the reflected path as directly observable [7], which can cause errors of several hundred meters in a deep urban canyon.

Low satellite visibility and poor geometry are other problems in urban GNSS positioning. Satellites are prone to be shadowed in urban areas due to high-rise buildings, and poor satellite visibility and lower number of visible satellites reduce both the availability and accuracy of GNSS positioning [12][13]. Moreover, only the satellites along track of the street are visible

This research was supported by the Ministry of Science and ICT, Korea, under the Information Technology Research Center support program (IITP-2020-2018-0-01423) supervised by the Institute for Information & Communications Technology Planning & Evaluation and Unmanned Vehicles Core Technology Research and Development Program through the National Research Foundation of Korea(NRF) and Unmanned Vehicle Advanced Research Center(UVARC) funded by the Ministry of Science and ICT, the Republic of Korea(No. 2020M3C1C1A01086407).

Y. Lee and B. Park are with the Department of Aerospace Engineering and Convergence Engineering for Intelligent Drone, Sejong University, Seoul, 05006, Republic of Korea (e-mail: byungwoon@sejong.ac.kr).

Y. Hwang is the KPS satellite navigation research center , Electronics Telecommunications Research Institute, Daejeon, 34129, Republic of Korea, (e-mail: ylhwang@etri.re.kr )

J. Ahn is the Autonomous unmanned vehicle research department, Electronics Telecommunications Research Institute, Daejeon, 34129, Republic of Korea, (e-mail: jyahn@etri.re.kr )

J. Seo are with the School of Integrated Technology, Yonsei University, Incheon 21983, Republic of Korea (e-mail:jiwon.seo@yonsei.ac.kr).

because of signal blocking by the buildings on both sides of the street, which causes the errors to be distributed in the form of an ellipse excessively elongated in the cross-track direction of the road. The influence of the geometrical distribution of satellites around the users [14] is considered a primary factor in lowering the GNSS positioning integrity [15]. Owing to very few redundant satellites and poor satellite geometry, the multipath errors of some satellites cannot be hidden in the position error; instead, they are more apparent than under the open sky, making the calculated position unreliable.

Recently, various studies have attempted to mitigate the multipath effect by using additional equipment such as fish-eye lens, inertial navigation system (INS), visual camera, and light detection and ranging (LiDAR). The technique using a fish-eye lens employs a camera synchronized with the GNSS and projects the satellites onto the photographed image to distinguish between the LOS and NLOS satellites [16]. This method has been suggested as a good option for independent sensors and for intuitively identifying NLOS satellites. Because fish-eye lenses are affected by various environmental and weather conditions [16], the methods involving omnidirectional infrared cameras that can be used even in dark environments such as night have recently been studied [17][18]. However, a GNSS receiver with an embedded fish-eye lens is difficult to be utilized universally due to difficulties in installation and infrared cameras cannot often distinguish buildings from the sky clearly.

The most traditional solution to mitigate multipath is GNSS/INS integration [10], which can provide continuous reliable positioning, GNSS-based heading information, and three-axis precision orientation under conditions considered challenging for GNSS [19]. As a more advanced type, the GNSS–inertial measurement unit (IMU)–camera combined system has also been proposed for better dynamic application of vision sensors to provide accurate relative positions [20][21]. Although the INS can provide an accurate relative position of a vehicle during a short time under normal conditions [22], the INS-integrated technique often fails to compute accurate solutions to overcome the multipath in deep urban areas if the vehicle has been driven with frequent stops without updating the GNSS solutions for a long time [12] [23].

LiDAR, a recently popular sensor for fusion with GNSS, can measure the distance to a target by illuminating it. It can be applied to a wide range of metallic and non-metallic targets [24]. Multipath on GNSS signals in urban environments is characterized with the help of environmental features extracted by the LiDAR [25]. A LiDAR-based perception method to exclude the NLOS receptions caused by moving objects in road transportation has been recently studied [26]. The sensors, which are integrated into the navigation systems employing GNSS, are able to compute only the relative position; therefore, it is necessary to apply the absolute position [26][27] provided by GNSS before time-updating. The accuracy of the integrated navigation system eventually depends on the GNSS; therefore, it is important to improve the absolute position accuracy of the GNSS in urban canyons.

Three-dimensional (3D) building information improves the accuracy of the GNSS absolute position in urban canyons; in addition, techniques such as shadow matching and ray tracing have been introduced. Shadow matching utilizes the 3D building information to check the visibility of each satellite based on the geometry formed by the user location, surrounding buildings, and GNSS signal direction[26][27][28]. The ray-tracing technique estimates and mitigates the pseudorange multipath errors by using all possible reflected rays from the satellite to the receiver based on the 3D building information [29]. Because ray tracing directly estimates the error of the NLOS signal, it has the advantage of mitigating the error without damaging the positioning availability.

The biggest challenge in shadow matching and ray tracing is the overwhelming computational load [10]. Because multipath errors differ depending on the signal reception position, both methods assign candidates near the initially calculated position and then exclude or compensate for the NLOS pseudorange observable for all the position candidates. Only the exact position among the candidates eventually enables the exclusion and compensation of the NLOS errors to compute the position accurately; this problem was called the chicken–egg problem by van Diggelen [10]. Thus, for integration with additional sensors, the position accuracy of the GNSS is vital for successful integration with information in urban areas.

There are many traditional techniques for mitigating multipath by using only GNSS measurements, namely, multi-constellation, measurement weighting, consistency checking, and averaging, as summarized in Table I. Multi-constellation is the simplest way to mitigate the multipath effect on the position results [30] and improve the availability, accuracy, and robustness in urban areas by increasing the number of available satellites [19]. The weighting technique based on the elevation angle or signal strength of each satellite is another widely used technique for mitigating the multipath effect [31]. However, the multi-constellation and/or weighting are/is effective in conveniently mitigating the influence of the multipath errors on the position rather than the multipath itself and are therefore ineffective when more than a few satellites are available.

Integrity control techniques, such as receiver autonomous integrity monitoring (RAIM) based on the measurement residuals, can also be an option to exclude the satellite with the largest residual error caused by the multipath error [32]. The conventional RAIM algorithm assumes many redundancies and no more than one failure; however, it is extremely difficult to apply to urban signal reception environments [14]. The monitoring code minus carrier (CMC) can be used as another metric to measure multipath variations [33]. Because this technique assumes that the multipath errors accumulated over a long period have a Gaussian distribution, it is necessary to solve the carrier ambiguity by averaging the CMC to calculate and correct the error. However, this assumption is acceptable only when long-term data have been accumulated, and the averaged value of multipath and NLOS errors in urban areas cannot be zero in urban areas.

TABLE I
MULTIPATH MITIGATION TECHNIQUES USING ONLY GNSS MEASUREMENT

| Classification | Technique | Accuracy | Availability |
|---|---|---|---|
| Multipath Mitigation | - Measurement Weighting<br>- Multi-Constellation | Partially increased | No loss of availability |
| Multipath Detection & Exclusion | - Consistency Checking | Increased | Loss of availability |
| Multipath Correction | - Averaging CMC<br>- CMC Variation | Increased | No loss of availability |

Despite the imperfectness of the assumption in the existing CMC method, our previous study focused on the availability of CMC to observe the multipath variation [34]. Once the initial multipath is computed correctly, the CMC variation enables the update of the multipath error, which directly mitigates the error regardless of LOS or NLOS. However, starting at a known coordinate is not a realistic condition, and the multipath estimation uncertainty can accumulate gradually when the estimation filter is processed for a long time.

To solve the inherent problem of the previous algorithm, this study proposes an automatic switching technique between general DGNSS and CMC-based multipath mitigation. The DGNSS residual-based validation test enables the switching algorithm to check the current signal environment and then determine a suitable positioning mode. Once the algorithm determines that the current GNSS observables are reliable, the DGNSS is activated for the current positioning mode. However, when the DGNSS position is considered no longer valid, the positioning mode is switched to the CMC-based multipath mitigation mode until the DGNSS residuals pass the validation test again.

The remainder of this paper is organized as follows. In Section II, a seamless multipath estimation methodology is described, considering deep urban applications. Section III presents the multipath mitigation and positioning method with its mode-switching technique. A field test was conducted in Teheran-ro, Seoul, Korea (Fig. 1), and the results are examined in Section IV. The discussion and conclusions are presented in Section V.

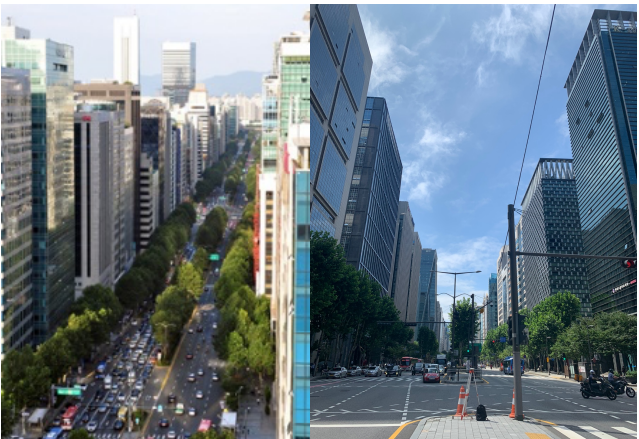

**Fig. 1.** Deep urban canyons in Teheran-ro (Seoul, South Korea).

## II. Multipath Estimation Methodology for Urban Environments

### A. Consecutive Multipath Estimation Using GNSS CMC Variation

A reflected signal can distort the GNSS correlation peak generated by the direct signal, which can cause multipath interference. The maximum error in pseudorange measurements is half of a ranging code chip, approximately 150 m, and the carrier phase error can reach a quarter of a wavelength, which is 4.76 cm for the GPS L1 frequency [35]. When an NLOS signal is received solely without a direct signal, the receiver mistakes the received signal as a direct signal. In this case, the reflected signal path length is induced into the range error; thus, there is no limit on the maximum error, which can be up to several kilometers. Except for the cycle slip case due to the transition between LOS and NLOS, the carrier-phase multipath error remains limited to half the cycle of the wavelength (modulo one carrier cycle) [7]. On the receiver side, the error-inducing mechanism, range of the LOS multipath interference, and NLOS reflected range error are very different from each other; however, the error terms are included in the measurements in the same format and cause positioning errors in the same manner. In this paper, we collectively call both errors induced by these two reflected signals, i.e., LOS and NLOS signals, as multipaths.

The multipath error ($M^i$) in the pseudorange measurement for the $i$-th satellite is included in the GNSS code observable ($\rho_f^i$) for frequency $f$, as described in (1). Similarly, the carrier phase multipath error ($m_f^i$) is included in (2), which is much smaller than $M^i$ for both LOS and NLOS cases.

The pseudorange code measurement and carrier phase measurement of the $i$-th satellite at time $t$ can be modeled as (1) and (2).

$$\rho_f^i(t) = d^i(t) + \left(B(t) - b^i(t)\right) + I_f^i(t) + T^i(t) + M_f^i + \epsilon_f^i \tag{1}$$

$$\phi_f^i(t) = d^i(t) + \left(B(t) - b^i(t)\right) - I_f^i(t) + T^i(t) + m_f^i + N_f^i \lambda_f + \varepsilon_f^i \tag{2}$$

where $d$ is the distance between the receiver and satellite, and $B$ and $b$ are the receiver and satellite clock errors, respectively. $I$ and $T$ denote the ionospheric and tropospheric errors, respectively. The measurement noise values of the pseudorange and carrier phases are represented by $\epsilon$ and $\varepsilon$. $N$ and $\lambda$ in the carrier phase modeling equation are the integer ambiguity and wavelength, respectively, for the frequency $f$. $M$ and $m$ represent the multipath errors included in the pseudorange and carrier phase measurements, respectively.

By linearly combining the code and phase measurements of the L1 and L2 frequencies in (1) and (2), the ionospheric errors in both the code and carrier measurements are removed, which results in the ionosphere-free linear combination equations (3) and (4).

$$\rho_{if}^i(t) = \frac{\rho_{L1}^i(t)\cdot\gamma - \rho_{L2}^i(t)}{\gamma - 1} = d^i(t) + \left(B(t) - b^i(t)\right) + T^i(t) +$$

$$M_{if}^i(t) + \epsilon_{if}^i(t) \qquad , \qquad (3)$$

$$\phi_{if}^i(t) = \frac{\phi_{L1}^i(t)\cdot\gamma - \phi_{L2}^i(t)}{\gamma-1} = d^i(t) + \left(B(t) - b^i(t)\right) + T^i(t) + m_{if}^i(t) + \epsilon^i(t) + \frac{\lambda_{L1} N_{L1}^i(t)\cdot\gamma - \lambda_{L2} N_{L2}^i(t)}{\gamma-1}, \qquad (4)$$

where $\gamma$ is the square of the L1/L2 frequency ratio, that is, $f_{L1}^2/f_{L2}^2$, and the subscript $if$ represents the ionosphere-free combined measurement.

If a rover's position at the initial time $t_0$ has been computed exactly, the initial value of the multipath error that corrupts the pseudorange code observable can be calculated as shown in (5). In the initial multipath error estimation, the distance ($\hat{d}^i$) from the rover's initial position to each satellite and the tropospheric error ($\hat{T}^i$) computed by the Saastamoinen model are used. The satellite clock bias $\hat{b}_{if}^i$ is corrected for the ionosphere-free combination by using the received navigation message. The rover's clock bias can be mitigated using $\hat{B}$ from the navigation solution, which may cause a bias in the multipath initial estimate because $\hat{B}$ cannot be accurate. However, the bias is common to all satellites; thus, it does not harm the rover's position accuracy.

$$M_{if}^i(t_0) \approx \rho_{if}^i(t_0) - \hat{d}^i(t_0) - \left(\hat{B}(t_0) - \hat{b}_{if}^i(t_0)\right) - \hat{T}^i(t_0) \quad (5)$$

Once the initial value of the pseudorange multipath has been estimated, its values can be updated using the CMC variation. The CMC process enables the removal of the geometry terms, that is, distance, clock offsets, and tropospheric error, which results in the modeling equation (6) comprising the code and carrier multipath. It should be noted that the code multipath error is incomparably much larger than that of the carrier phase. Thus, a simpler equation is obtained, as shown in (7), which is mainly composed of the code multipath and ambiguity of the carrier phase.

$$CMC_{if}^i(t) = \rho_{if}^i(t) - \phi_{if}^i(t)$$
$$= M_{if}^i(t) - m_{if}^i(t) - N_{if}^i(t)\lambda_{if} + \epsilon_{if}^i(t) - \varepsilon_{if}^i(t) \qquad (6)$$

$$CMC_{if}^i(t) \approx M_{if}^i(t) - N_{if}^i(t)\lambda_{if} + \epsilon_{if}^i(t) - \varepsilon_{if}^i(t) \qquad (7)$$

If the receiver has not lost its continuous carrier tracking while maneuvering in the urban canyons, the cycle of the carrier would not slip, and the ambiguity term remains constant. Because the time difference of the carrier ambiguity, $\Delta N_{if}^i(t)\lambda_{if}$, is zero under the condition that no cycle slip occurs in the $i$-th satellite carrier observation, the variation in the code multipath, $\Delta M_{if}$, is almost the same as the CMC variation, as shown in (8).

$$\Delta CMC_{if}^i(t) \approx \Delta M_{if}(t) \qquad (8)$$

We can calculate the code multipath error for the $i$-th satellite at time $t$ by accumulating the CMC variation from the initial value obtained from (5). Because we suggest using the ionosphere-free combination measurements, unlike other studies, the suggested multipath calculation of (9) is free from ionospheric divergence even if it is accumulated for a long time.

$$M_{if}^i(t) = M_{if}^i(t_0) + \sum_{k=t_0}^{t} \Delta CMC_{if}^i(k) \qquad (9)$$

### B. Compensation for Time Offset Between Constellations

Another serious problem that hinders the availability of urban GNSS navigation as much as multipath errors is the low visibility of GNSS satellites. Because skyscrapers in a city diminish the visibility of GPS satellites, multi-constellation GNSS solutions are essential for the continuous navigation of vehicles in deep urban areas. Navigation availability increases as more satellites are visible, but this increment is not perfectly proportional to the number of visible satellites. All GNSSs inherently depend on precise timekeeping; therefore, each ground segment dedicates considerable effort toward maintaining a highly stable atomic timescale. Nevertheless, clock differences between the constellations continue to exist among the timescales at the level of tens or hundreds of nanoseconds [36]; thus, at least two more satellites should be added to improve the availability of multi-constellation GNSS positioning. For example, when a user cannot obtain the position solution owing to the visibility of three GPS satellites, an additional satellite in GLONASS or other constellations does not help the user, because the dimension of the state has increased from four to five owing to the inter-system bias.

Incorporating two additional satellites per constellation remains a challenge in deep urban areas. To increase the accuracy by mitigating the multipath errors and availability by adding a constellation, this study proposes the inclusion of inter-system bias into the initial multipath estimates. If the initial multipath and its variation are estimated based on the GPS system clock, multi-constellation positioning is possible with only four satellites, regardless of the number of constellations used for the positioning. As shown in (10), other systems do not use the same time reference as GPS, and thus a time difference between the other GNSS and GPS, denoted as $\delta B_{GNSS}$ in (11), arises.

$$\rho_{if}^I(t) = d^I(t) + \left(B_{GPS}(t) - b^I(t)\right) + T^I(t) + M_{if}^I(t) + \epsilon_{if}^I(t) \Big\}$$
$$\rho_{if}^J(t) = d^J(t) + \left(B_{GNSS}(t) - b^J(t)\right) + T^J(t) + M_{if}^J(t) + \epsilon_{if}^J(t) \Big\}$$
$$, \qquad (10)$$

where the superscripts $I$ and $J$ refer to the GPS and non-GPS GNSS satellites, respectively.

$$B_{GNSS}(t) = B_{GPS}(t) + \delta B_{GNSS}(t) \qquad (11)$$

Receivers using measurements from two or more systems need to cope with this time offset; thus, the minimum number of visible satellites should be increased by the number of added constellations to compensate for the difference in clock error [37]. For this multi-constellation GNSS solution, the navigation matrix $H_{GNSS}$ should be described as shown in (12), where $\vec{e}$ is the LOS unit vector from the user to each GNSS satellite.

$$H_{GNSS} = \begin{bmatrix} \vec{e}_{GPS}^{\,i} & -1 & 0 & 0 \\ \vdots & \vdots & \vdots & \vdots \\ \vec{e}_{GLONASS}^{\,i} & 0 & -1 & 0 \\ \vdots & \vdots & \vdots & \vdots \\ \vec{e}_{Beidou}^{\,i} & 0 & 0 & -1 \\ \vdots & \vdots & \vdots & \vdots \end{bmatrix} \quad (12)$$

In the deep urban areas targeted by this study, extremely few GNSS satellites are visible, even when all the available constellations are used. Regardless of its accuracy, multipath estimation is useless if the GNSS position cannot be solved. Therefore, if $\delta B_{GNSS}$ is included in the multipath estimation process, multi-constellation navigation availability is improved, as is its accuracy due to multipath mitigation. To utilize this algorithm, $\delta B_{GNSS}$ should be added to $\hat{M}_{if}^j$, and the multipath estimates of GNSSs other than GPS with reference to GPSTime, $\hat{M}_{if|GPST}^j$, can be obtained as shown in (13).

$$\hat{M}_{if|GPST}^j(t) = \hat{M}_{if}^j(t) + \delta B_{GNSS}(t) \quad (13)$$

The multipath estimate at the first epoch referenced to GPSTime, $\hat{M}_{if|GPST}^j(t_0)$, is easily calculated by inserting $B_{GPS}(t)$ into (5) instead of $B(t)$ of each constellation system, as described in (14).

$$M_{if|GPST}^j(t_0) \approx \rho_{if}^j(t_0) - \hat{d}^j(t_0) - \left(\hat{B}_{GPS}(t_0) - \hat{b}_{if}^j(t_0)\right) - \hat{T}^j(t_0) \quad (14)$$

Note that because each GNSS maintains a highly stable timescale, $\delta B_{GNSS}(t)$ changes very slowly, typically in the order of femtoseconds per second [37], and we can assume that the inter-system bias between $t_0$ and $t$ is constant [38]. Although the variation in $\delta B_{GNSS}$ from $t_0$ to $t$ is negligible for maneuvering general vehicles, CMC is also referenced by GPSTime to avoid any unexpected divergence. Now, the multipath estimate equation of the non-GPS satellite at time $t$ is changed from (9) to (15).

$$\hat{M}_{if|GPST}^j(t) = M_{if|GPST}^j(t_0) + \sum_{k=t_0}^{t} \Delta CMC_{if|GPST}^j(k) \quad (15)$$

When a receiver applies the multipath estimates of the $i$-th GPS and $j$-th non-GPS satellites to its measurements simultaneously, it can compensate for the inter-system bias while mitigating the multipath errors, as described in (16).

$$\left.\begin{array}{l} \rho_{if}^i(t) - \hat{M}_{if}^i(t) - \hat{T}^i + \hat{b}^i \approx d^i(t) + B_{GPS}(t) \\ \rho_{if}^j(t) - \hat{M}_{if|GPST}^j(t) - \hat{T}^j + \hat{b}^j \approx d^j(t) + B_{GPS}(t) \end{array}\right\} \quad (16)$$

All the pseudoranges in (16), unlike those in (10), are synchronized to GPSTime, and it enables the shrinking of the observation matrix of (12) to that of (17). The GPSTime-synchronized navigation equation including the observation matrix of (17) requires only four visible satellites even when using a multi-constellation, which is expected to enlarge the navigation availability and is therefore suitable for urban areas.

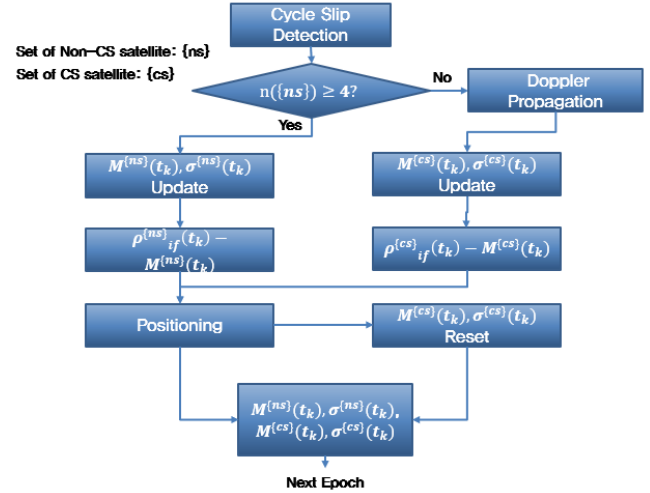

**Fig. 2.** Consecutive estimation process of CMC-based multipath and its standard deviation.

$$H_{GNSS|GPST} = \begin{bmatrix} e_{GPS}^i & -1 \\ \vdots & \vdots \\ e_{GLONASS}^j & -1 \\ \vdots & \vdots \\ e_{Beidou}^k & -1 \\ \vdots & \vdots \end{bmatrix} \quad (17)$$

### C. Consecutive Estimation Process of CMC-Based Multipath

The multipath estimation algorithm described above is effective under the assumption that the integer ambiguity remains constant. To check if the algorithm can be continuously available for current measurements, a cycle-slip detection technique was applied to two consecutive carrier phases. The time difference of the geometry-free combination of the dual-frequency carrier measurements was used as a cycle-slip detection metric, as shown in (18).

$$\Delta I_{L1}^i(t) = \frac{\Delta \phi_{L1}^i(t) - \Delta \phi_{L2}^i(t)}{\gamma - 1} + \frac{[\Delta N_{L2} \lambda_{L2}(t) - \Delta N_{L1} \lambda_{L1}(t)]}{\gamma - 1} \quad (18)$$

Because ionospheric variation in time is less than 2 cm/s in the mid-latitude region [39], the metric in (18) being larger than the normal value means that the time difference of the integers, $\Delta N_{L1}$ or $\Delta N_{L2}$, is not zero due to the cycle slip. Fig. 2 describes the process of multipath estimation when cycle slips are found in the current satellite measurements.

Once a cycle-slip alarm is flagged for a satellite at time $t_k$, the consecutive multipath estimation algorithm is no longer valid for the detected satellite {cs}. Consecutive multipath estimates for the non-slipped satellites {ns} are valid; thus, the position solution at time $t_k$ is still reliable owing to the exclusion of the slipped satellite from the solution, provided the number of non-slipped satellites is four or more. Because the position obtained after mitigating the multipath and ionospheric errors for the non-slipped satellites by using (3) and (16) is reliable, the multipath estimate of the slipped satellite $\widehat{M}_{if|GPST}^{\{cs\}}$ can be initialized again based on the obtained position using (14).

Recalling the estimate equation of (15), the error variance of the CMC-based multipath estimates, $\left(\sigma_{\widehat{M}}^i\right)^2$, is the sum of the first estimate's variance and accumulated variance of the CMC time difference, as described in (19). From (6), the CMC includes code noise as well as multipath error in the pseudorange. The code noise and multipath are removed together when $\widehat{M}_{if}^i$ estimated by the accumulated CMC is subtracted according to (16), and the variation in the carrier phase residual error is accumulated. The time difference of the carrier phase measurement has a millimeter-level residual, which is known to be more accurate than the centimeter-level residual of Doppler [40][41]. Here, $\sigma_{\Delta CMC_{if}}^i$ was set to 3 cm/s owing to inflation due to the dual frequency combination. Therefore, the uncertainty increases with time after the multipath estimation is initialized.

$$(\sigma_{\widehat{M}_{if}}^i(t))^2 = (\sigma_{\widehat{M}_{if}}^i(t_0))^2 + \sum_{k=t_0+1}^{t}(\sigma_{\Delta CMC_{if}}^i(k)\Delta t)^2, \quad (19)$$

where $\sigma$ is the standard deviation (std) of each error denoted by the subscript, and $\Delta t$ is the sampling time.

If the number of available satellites is less than four due to cycle slip and low-visibility, dual-frequency carrier phase measurements should be replaced by ionosphere-free (*if*) combined Doppler measurements in (20) to extend the *if* carrier measurement at the previous epoch, as shown in (21). When the Doppler measurements were used, $\sigma_{\Delta CMC_{if}}^i$ was inflated to 30 cm/s.

$$d\rho_{if}^{\{cs\}}(t) = \frac{d\rho_{L1}^{\{cs\}}(t)\cdot\gamma - d\rho_{L2}^{\{cs\}}(t)}{\gamma - 1} \quad (20)$$

$$\phi_{if}^{\{cs\}}(t) = \phi_{if}^{\{cs\}}(t-1) + \frac{d\rho_{if}^{\{cs\}}(t) + d\rho_{if}^{\{cs\}}(t-1)}{2} \quad (21)$$

After solving the rover's position at time $t$, the multipath estimates and their uncertainties for the slipped satellites, $\widehat{M}_{if|GPST}^{\{cs\}}$ and $\sigma_{\widehat{M}_{if}}^{\{cs\}}$, are initialized.

## III. Mode-Switching and Positioning Methodology

### A. DGNSS Validation Test

To implement the consecutive multipath estimation process, two problems should be solved in advance: how to determine the initial position and whether the initial position is reliable. Differential GNSS (DGNSS), which is accurate and reliable code-based GNSS technique, has an accuracy of 1 m. However, there is no way to confirm whether the currently obtained position is reliable without the addition of a camera or 3D map. Instead of adding extra sensors or geographical information, a DGNSS validation test based on residuals was utilized in this study.

The least-square residual vector ($\hat{v}$) for the DGNSS solution $\widehat{X}_{DGNSS}$ can be expressed as (22), and the corresponding covariance matrix for the DGNSS covariance matrix $P_{\widehat{X}_{DGNSS}}$ is described in (23).

$$\hat{v} = z - H\widehat{X}_{DGNSS}, \quad (22)$$

$$P_{\hat{v}} = R - HP_{\widehat{X}_{DGNSS}}H^T, \quad (23)$$

where $z$ represents the measurement vector of the pseudorange after applying the pseudo-range correction (PRC). Under the open sky, most bias errors such as satellite-related and atmospheric errors are mitigated by feeding the PRC to the code observables. The DGNSS residual errors are close to white noise, with deviations dependent on the elevation angle [42]. Assuming that there is no correlation between the satellites, the measurement noise covariance matrix $R$ is given by (24).

$$R = diag\left(\sigma_{DGNSS}^i{}^2\right), \quad (24)$$

where $\sigma_{DGNSS}^i$ is the std of $z$ in (22), which is the multipath sum of the rover ($rv$) and code noise of both the reference station ($rs$) and $rv$, as shown in (25). The multipath ($\sigma_m$) and noise error ($\sigma_n$) models defined for the normal conditions in RTCA standard [43] were used.

$$\sigma_{DGNSS}^i = \sqrt{\sigma_m^2(el^i) + \sigma_{n,rv}^2(el^i) + \sigma_{n,rs}^2(el^i)}, \quad (25)$$

where $\begin{cases} \sigma_m(el) = 0.15 + 0.43e^{-el/6.9°} \\ \sigma_n(el) = 0.13 + 0.53e^{-el/7.5°} \end{cases}$

When the residuals are sufficiently normal to make the assumption of the zero-mean Gaussian distribution valid, the weighted square sum of error (WSSE) of the DGNSS residuals obeys the chi-square distribution ($\chi^2$). However, if any instance of $\hat{v}$ contains a mix of biases, mostly due to the multipath error in urban areas, the WSSE becomes too large to follow the $\chi^2$ distribution. Thus, *WSSE* in (26) can represent the multipath detection function and it can be used to determine the current rover's signal reception environment.

$$WSSE = \hat{v}^T P_{\hat{v}}^{-1} \hat{v} \quad (26)$$

According to the Neyman–Pearson criterion, when the false alarm rate $\alpha$ is determined, the threshold $T_d$ is determined by solving the equation $P(WSSE > T_d) = \alpha$ [44]. Here, we set the false-alarm rate to 0.01%.

Therefore, it can be considered that all the GNSS signals have been received in suburban or open-sky areas if *WSSE* passes the chi-square test in (26). Because the current DGNSS solution is reliable, all multipath estimates can be initialized, and the CMC-based multipath process is ready to start.

Conversely, if *WSSE* is greater than $T_d$, some of the visible satellites are suspected to be severely corrupted due to multipath in the urban area, and the DGNSS result is no longer reliable. It is difficult to apply the widely used RAIM method because the number of satellites that are seriously affected by the multipath is unknown. However, the proposed algorithm can estimate the multipath errors continuously from the initial position regardless of the number of satellites that are corrupted by multipaths and the source of the multipath error (LOS or NLOS). Therefore, the rover can determine its accurate position by effectively mitigating all multipath errors, even in a deep urban area.

$$\begin{cases} WSSE > T_d & \text{severe multipath (SM)} \\ WSSE \leq T_d & \text{multipath} - \text{free (MF)} \end{cases} \tag{27}$$

Even without any camera or building information, the rover can determine whether there are any obstacles that cause multipath based on the signal reception decision criterion of (27). Because the final positioning algorithm depends on the signal reception environment, it is used as the positioning mode-switching criterion, which is explained in the next section.

### B. Positioning Mode-Switching Algorithm

Once the signal reception decision criterion determines that the rover is currently under a "severe multipath (SM) environment," the DGNSS position is no longer valid. The current position should be computed after mitigating the multipath by using the suggested estimation algorithm until the environment is changed to "multipath-free (MF)" because of a change in the rover's maneuvering or satellite geometry. Although the recursively estimated multipath improves the position accuracy of urban users using (17) and (18), the uncertainty gradually increases owing to the accumulation of the carrier phase time difference or Doppler, as shown in (19). The uncertainty does not increase with the same magnitude because the initialization time is different for each satellite; however, the noise covariance model for the measurements in the SM environment in (28), $R_{z_{SM}}$, increases in proportion to time. Consequently, the SM mode positioning error covariance $P_{\hat{\vec{X}}_{SM}}$ inevitably increases. Therefore, it is risky to run the multipath estimation filter over a long period from the initial multipath-fix time.

$$R_{z_{SM}}(t) = diag\left\{ \left(\sigma_{\hat{M}_{lf}}^i(t)\right)^2 + \left(\sigma_{n,rs}^i(t)\right)^2 \right\} \tag{28}$$

The moment when the multipath environment is switched from SM to MF is a good opportunity to renew the initialization time. Unlike the SM condition, the DGNSS position $\vec{X}_{DGNSS}$ and multipath-mitigated position $\hat{\vec{X}}_{SM}$ can be both valid in the MF environment. By combining two valid positions, $\vec{X}_{DGNSS}$ and $\hat{\vec{X}}_{SM}$, based on their estimated error covariances $P_{\vec{X}_{DGNSS}}$ and $P_{\hat{\vec{X}}_{SM}}$, the optimal position for the MF mode can be estimated as (29).

$$\left.\begin{aligned} \hat{\vec{X}}_{MF}(t) &= P_{\hat{\vec{X}}_{MF}}(t)\left(P_{\vec{X}_{DGNSS}}^{-1}(t)\vec{X}_{DGNSS}(t) + P_{\hat{\vec{X}}_{SM}}^{-1}(t)\hat{\vec{X}}_{SM}(t)\right) \\ P_{\hat{\vec{X}}_{MF}}(t) &= \left(P_{\vec{X}_{DGNSS}}^{-1}(t) + P_{\hat{\vec{X}}_{SM}}^{-1}(t)\right)^{-1} \end{aligned}\right\} \tag{29}$$

## IV. FIELD TEST IN DEEP URBAN AREA AND RESULTS

### A. Field Test Construction

To verify the applicability of the algorithm to actual deep urban areas, a field test was performed along Teheran-ro, Seoul, Korea, which is one of the streets with the poorest GNSS visibility and signal reception globally. According to the analysis based on the 3D building information in this area, only 2.5 GPS satellites are visible on average and GPS-only positioning is available for 20% of the day [45]. Although this area is within the "GNSS satellite hotspot," [46] where all constellation signals in operation are available, only 4 satellites are visible despite using all the constellations, and the multi-GNSS position is not available for approximately 10% of the day [47]. The low visibility and availability result in positioning vulnerable to multipath and poor accuracy. The average position error RMS in this area was reported to be 55.6 m due to the significant effect of multipath [48]. The dynamic test trajectory, which included Teheran-ro, is shown in Fig. 3. The starting point was a relatively suburban area, but most of the driving was on the road in the middle of a deep urban area. There was a bridge at the border between the suburban area near the starting point and the deep urban area, in which all signals were blocked before entering the urban area.

For the dynamic field test, a vehicle equipped with a GNSS receiver and reference system was used, as shown in Fig. 4. A NovAtel FlexPak6 GNSS receiver was used to receive the GPS/GLONASS/BeiDou dual-frequency pseudorange and carrier phase measurements, and NovAtel SPAN-CPT provided a continuous 3D true reference trajectory even when signal reception was briefly compromised. The GNSS signal recorder LabSat 3 was also mounted with other devices to acquire the position results to be compared by reradiating the same signals to other modes of the receiver.

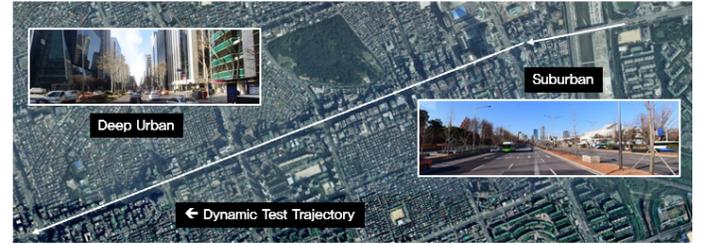

**Fig. 3.** Dynamic test trajectory.

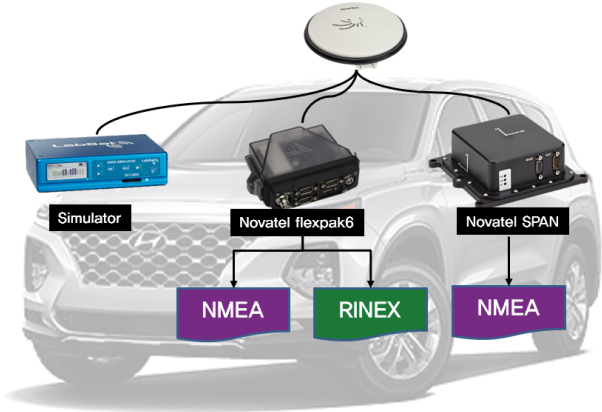

**Fig. 4.** Dynamic test configuration.

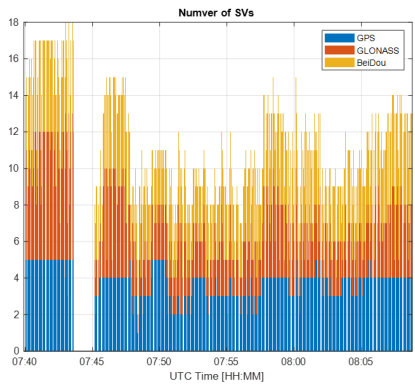

**Fig. 5.** Satellite visibility during dynamic test.

The test was carried out for approximately half an hour from 7:40 to 8:10 UTC on September 14, 2019. Fig. 5 shows the satellite visibility during the dynamic test. The average number of available GPS satellites was 3.74 during the dynamic test, and the average number of available satellites was 10.8 even after adding two more constellations, GLONASS and BeiDou. Because at least six satellites are required for three-constellation positioning, the conventional technique cannot provide the positions for approximately 96.6% of the test duration.

### B. Field Test Results

The multipath error for all visible satellites estimated on Teheran-ro according to the algorithm described in Section II is shown in Fig. 6. One notable aspect is that there are two groups of estimates with values centered at 0 and 30 m. The estimates mainly clustered at 30 m are the multipath errors of the BeiDou satellite, which were estimated by referring to the GPSTime and not the BeiDou time according to (14). This means that the clock offset between GPS and BeiDou was approximately 30 m (≈100 ns) during the dynamic test, as shown in Fig. 7. The clock offset between GPS and BeiDou computed under the open sky at the Sejong University reference station (Fig. 7, left) was similar to the estimated bias calculated (Fig. 7, right) using the suggested technique during the dynamic test in the deep urban area.

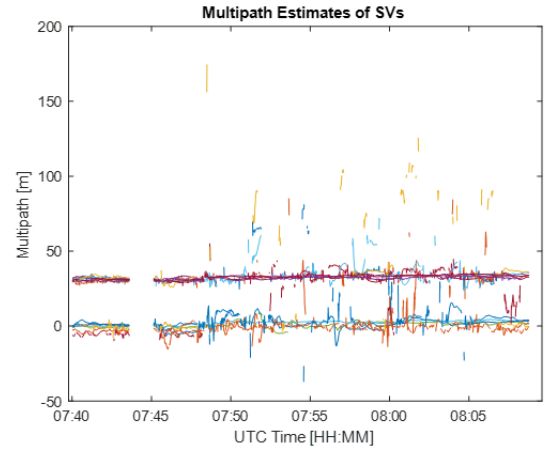

**Fig. 6.** Estimated multipath of multi-constellation satellites.

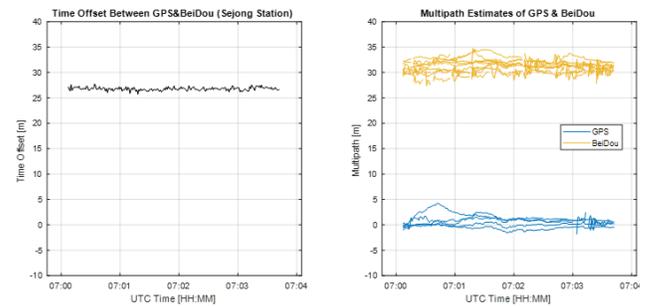

**Fig. 7.** Estimated time offset between GPS and BeiDou.

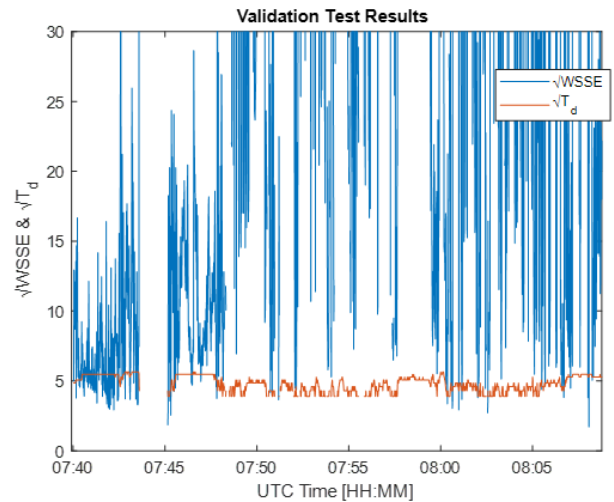

**Fig. 8.** DGNSS residual validation test results.

Considering the bias due to the clock offset between the systems, a large number of real multipath errors were estimated to be less than 50 m, but some were larger than 150 m, which confirms that NLOS multipaths were included in the measurements. Thus, the proposed algorithm can estimate the multipath error regardless of whether it is LOS or NLOS. The estimates are highly discrete because the vehicle movement constantly changes the satellite geometry and surrounding buildings. In this study, in order to evaluate how well the multipath estimates mitigate the actual errors, all visible

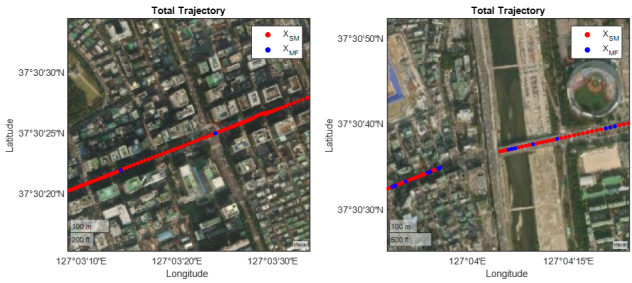

**Fig. 9.** SM and MF points in urban area.

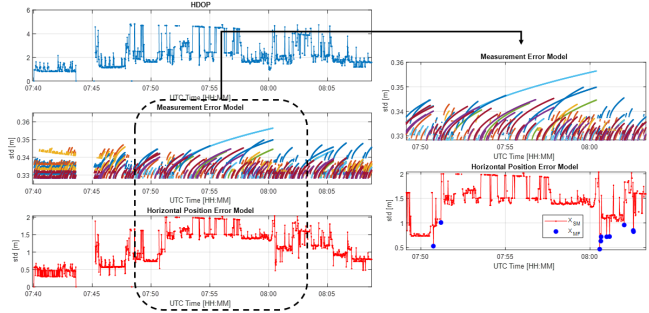

**Fig.10.** HDOP and error model std variation in measurement and position domains.

satellites were used for positioning without applying a satellite exclusion technique such as RAIM.

The rover can detect if the vehicle is currently driving in the SM or MF environment based on the signal reception decision criterion of (26) without additional sensors or information. According to our test results in Fig. 8, the DGNSS positions were acceptable for only 6% of the dynamic test. The points judged as MF were mostly located at the starting and ending points in relatively suburban areas. In urban areas from 7:45 to 8:08 UTC, the $WSSE$s at most points were calculated to be larger than $T_d$; therefore, the DGNSS results were mostly unreliable. However, the DGNSS results at 31 points among them were determined to be calculated in MF environments, which are marked on the maps in Fig. 9. Given that these points are all located near intersection points, where relatively good satellite visibility is available, this criterion is proven valid for determining the signal reception environment.

These DGNSS-valid points within urban canyons prevent the divergence of positioning error by stopping the uncertainty accumulation in each measurement considered in (19). The functionalities of these points in the measurement and position domains are presented in Fig. 10, especially in the dotted box magnified for the duration from 07:48 to 08:03. As shown in the map of Fig. 9, the horizontal dilution of precision (HDOP) at those points was temporarily very low when compared with the other points because the criterion of (26) identified sites that provided good visibility. At these points, $\hat{X}_{MF}$ is computed by optimizing $\vec{X}_{DGNSS}$ and $\hat{X}_{SM}$, and its error covariance $P_{\hat{X}_{MF}}$ is significantly reduced from $P_{\hat{X}_{SM}}$ accumulated until the previous epoch, as shown in the third row. Based on the reliable MF position $\hat{X}_{MF}$, all the multipath estimates $\hat{M}_{i f|GPST}^{j}$ were initialized, and their stds were reduced to the initial values.

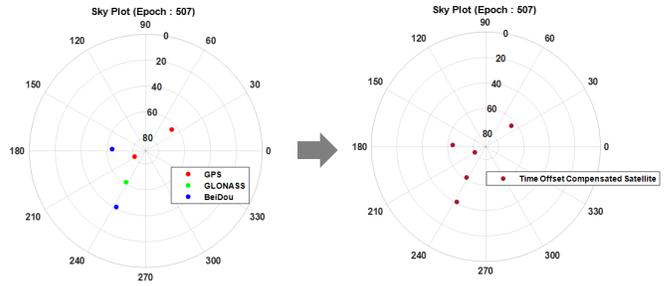

**Fig. 11.** Multi-constellation satellite geometry (left: conventional method, right: GPSTime-synchronized method)

Because there was no valid DGNSS result from 7:51:11 to 8:00:34 UTC, the multipath error was continuously estimated without initialization, and the uncertainty for all the satellites was accumulated, as shown in the magnified box of Fig. 10.

The DGNSS result was initially determined to be valid at 7:51:11 UTC, and the multipath estimates for all the satellites were initialized and fixed based on the reliable position that resulted in reduced std. The variation in the position error modeling was similar to that of the measurement. Because the combined position accuracy of the DGNSS and SM position is expected to be accurate and reliable, these points in the deep urban areas greatly contribute to preventing the position error accumulation and divergence to an incorrect position.The multipath estimation based on the synchronization to GPSTime greatly contributed to improving the availability. Fig. 11 shows a skyplot snapshot at the 507th epoch, when five satellites in three constellations, namely, GPS, GLONASS, and BeiDou, were visible. Because GLONASS and BeiDou are operated based on their own clock systems, two more state variables, i.e., a total of six variables, are necessary for the general multi-GNSS positioning techniques. The conventional method definitely could not compute the position, but the algorithm in this study could solve it with one redundancy because it required only four variables in the solution state owing to the elimination of the inter-system clock offset. This example clearly demonstrates that the proposed method is suitable for improving position availability in deep urban areas where satellite visibility is very limited.

Doppler-based multipath estimation propagation complements CMC-based estimation under the condition of low visibility of satellites. Because the time difference of the carrier observable is utilized for the CMC-based estimation method, the condition that valid carrier phases must be consecutively measured at the current epoch as well as the previous epoch and the condition of no cycle slip must be satisfied. Fig. 12 shows the point at which the multipath was estimated using the Doppler measurement during the dynamic test and its position error. As shown in Fig. 12, valid consecutive positions were provided by the Doppler propagation, even when the carrier phases of only three satellites were available due to cycle slip.

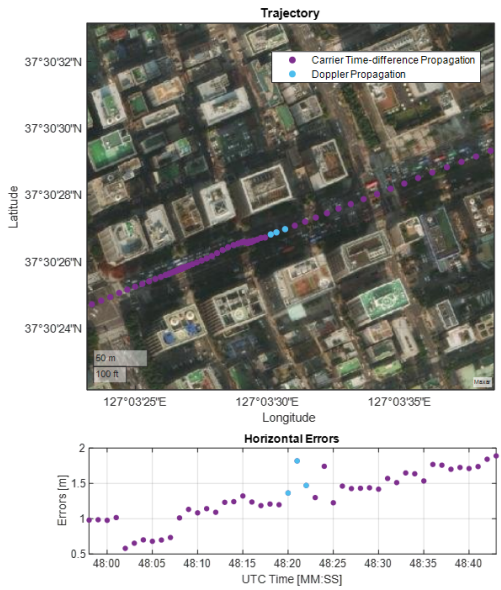

**Fig. 12.** Position complement by Doppler-based method.

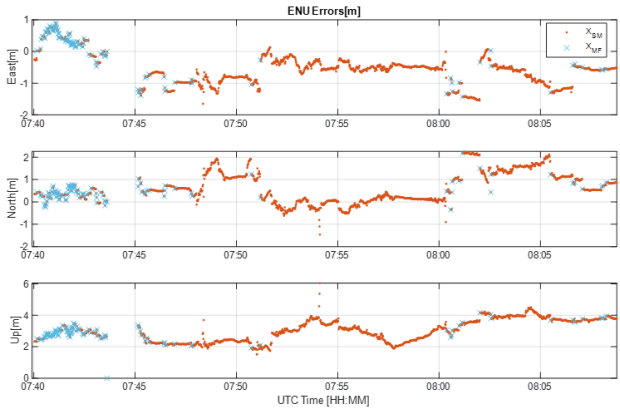

**Fig. 13.** DGNSS residual validation results.

As described above, the signal reception environment of the rover at each time was identified based on the DGNSS residual test; then, the multi-constellation GNSS multipath errors were mitigated by the GPSTime-synchronized CMC-based consecutive estimation technique for all the tests, and Doppler measurements were used as complementary measurements at the sites with extremely poor availability.

The multipath-mitigated positions of $\hat{X}_{SM}$ were combined with $\hat{X}_{DGNSS}$ to determine the final positions in the urban areas. The east-north-up (ENU) error at each time point was computed based on the true trajectory from the SPAN, as shown in Fig. 13. The positions were successfully computed during the entire 30-min dynamic test in Tehran-ro, and the root mean square (RMS) of the horizontal positioning error was computed as 1.2 m. The 95% cumulative error was 2.3 m, and the maximum error was only 2.7 m.

These results are superior to those of other conventional methods in terms of accuracy and availability. The GNSS signals recorded by LabSat-3 during the dynamic test were re-radiated to the NovAtel FlexPak6 in various modes, and real-time DGNSS, satellite-based augmentation system (SBAS), and RTK were replayed using the method presented in our previous study [49][50]. The trajectory of the suggested method along with those of RTK, DGNSS, and SBAS for the NovAtel receiver is shown in Fig. 14. The availability of the suggested method was 100%, whereas the DGNSS technique could calculate the position for 64.8% of the test section.

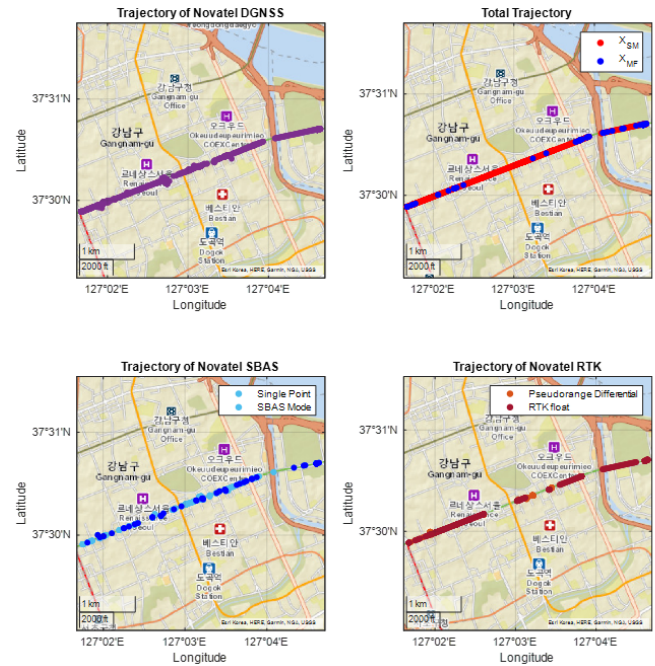

**Fig. 14.** Position availability comparison for various modes.

In addition to the availability, the RMS error of the proposed method was only approximately 1/10 of that of the DGNSS horizontal results, and the maximum error of over 100 m was reduced to less than 3 m as shown in Fig. 15 and Table II. Because the SBAS correction was applied to only GPS satellites, the receiver in the SBAS mode was able to calculate the position for only 4.2% of the test section. The overall accuracy, i.e., the horizontal RMS, was 11.3 m, as in the case of DGNSS, which has the same code differential positioning. However, the maximum error and 95% error of SBAS are statistically better than those of DGNSS because SBAS positioning is possible at points with better visibility when compared with multi-constellation DGNSS. When the receiver was in the RTK mode, the position was computed in 52.3% of the session, but no positions were computed in the RTK fixed mode whereas the positions were computed for 40% of the session in the RTK float mode. Despite the computation in the RTK float mode, the performance was assessed in terms of the horizontal error RMS instead of the commonly expected cm or dm accuracy; the error was computed to be 14 m, which was larger than the code-based results. The maximum error also reached 115 m horizontally and 216 m vertically; thus, the technique could not take advantage of the carrier-based positioning at all.

TABLE II
STATISTICAL RESULTS OF DYNAMIC TEST

| Statistical Results | | RMS | 95% | Max | Availability |
|---|---|---|---|---|---|
| Multipath-Free Position | Hor | 1.2m | 2.3m | 2.7m | 100% |
| | Ver | 3.0m | 4.0m | 6.0m | |
| Flexpak6 DGNSS | Hor | 11.1m | 15.6m | 101.0m | 64.8% |
| | Ver | 3.9m | 13.7m | 127.6m | |
| Flexpak6 SBAS | Hor | 11.3m | 25.7m | 62.6m | 4.2% |
| | Ver | 4.0m | 8.6m | 17.13m | |
| Flexpak6 RTK | Hor | 14.3m | 37.1m | 115.0m | 0.0%(fixed) |
| | Ver | 11.5m | 23.4m | 216.6m | 40.1%(float) |

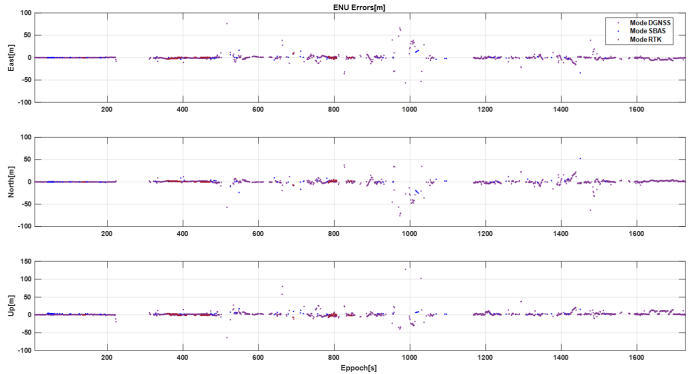

**Fig. 15.** Position accuracy comparison for various modes.

## V. CONCLUSIONS

This paper introduced an effective algorithm for mitigating severe multipath to determine the DGNSS positioning accuracy level consistently in deep urban areas without additional sensors or information other than GNSS. The existing methods selectively enhanced either the accuracy or availability, but the suggested method could improve both performance metrics. This method was able to estimate both LOS and NLOS without any algorithm variant, which also distinguishes it from other studies.

To validate the suggested algorithm and demonstrate its performance in urban canyons, we conducted a dynamic test in Teheran-ro, which is known for low visibility of satellites. The half-hour driving test results showed that the proposed algorithm could provide the rover's real-time results for 100% of the session, whereas the conventional method of the commercial receiver calculated the positions for only 64% of the session. The horizontal error RMS of the commercial receiver was 11.1 m, which was reduced to 1.2 m by applying the suggested algorithm to the receiver. From this result, we confirmed that the recognition of the signal reception environment and estimation of the GNSS multipath was valid and adequate to enable 1 m accuracy and 100% availability in a deep urban area.

Because this method does not require any prior information before deployment, it is expected that consistent position performance will be achieved in cities other than Seoul. This technique is very easy to implement when only the receiver provides dual frequency measurements; therefore, it is expected that it will be widely used not only in automobiles but also in various future intelligent transportation systems, including smart mobility based on electric bicycles, scooters, and drones. A precise position can be obtained after the vehicle obtains a reliable initial position such as a valid DGNSS. Our future research will focus on finding the initial position with sufficient accuracy to start this algorithm quickly in the middle of an urban canyon.

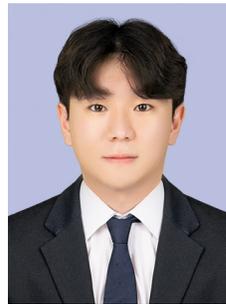

**Yongjun Lee** (Graduate Student Member, IEEE) is a Ph.D. student at the Department of Aerospace Engineering and Department of Convergence Engineering for Intelligent Drone in Sejong University in Republic of Korea. He received the B.S. and M.S. degrees from Sejong University. His research interests include GNSS, multipath, urban positioning, and machine learning.

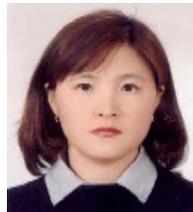

**Yoola Hwang** received her BS degree in Yonsei University, Seoul, Korea. She received her MS degree in aeronautics and astronautics from Purdue University, W-Lafayette and the PhD degree in aerospace engineering sciences from University of Colorado, Boulder, USA, respectively. She joined ETRI in 2004, where she developed the flight dynamics of ground control system of the LEO and GEO satellites and researched GNSS module of UAV. Recently she is involving in Korean Positioning System (KPS) project.

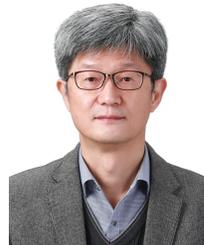

**Jae Young Ahn** received the BS, MS, and PhD degrees in engineering from Yonsei University, Seoul, Korea, in 1983, 1985, and 1989, respectively.
Since 1989, he has been with the ETRI, Taejon, Korea. From 1989 to 2016, he was involved in the development of satellite communications systems, wireless LAN technologies, and radio transmission schemes for mobile communications in ETRI. Since 2017, he has been with autonomous unmanned vehicle research department, ETRI, where he is now an assistant vice president. His current research interests include the autonomous operation of unmanned vehicles, the UAM avionics and CNSi infrastructure, and the sUAS and UAM traffic management.

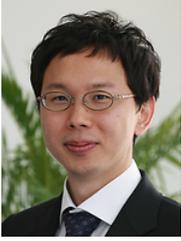

**Jiwon Seo** received the B.S. degree in mechanical engineering (division of aerospace engineering) in 2002 from the Korea Advanced Institute of Science and Technology, Daejeon, Korea, and the M.S. degree in aeronautics and astronautics in 2004, the M.S. degree in electrical engineering in 2008, and the Ph.D. degree in aeronautics and astronautics in 2010 from Stanford University, Stanford, CA, USA. He is currently an Associate Professor with the School of Integrated Technology, Yonsei University, Incheon, Korea. His research interests include complementary positioning, navigation, and timing systems, Global Navigation Satellite System (GNSS) anti-jamming technologies, and ionospheric effects on GNSS. Prof. Seo is also a Member of the International Advisory Council of the Resilient Navigation and Timing Foundation, Alexandria, VA, USA, and a Member of several advisory committees of the Ministry of Oceans and Fisheries and the Ministry of Land, Infrastructure and Transport, Republic of Korea.

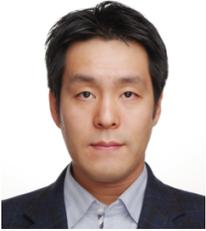

**Byungwoon Park** (Member, IEEE) received his B.S., M.S., and Ph.D. degrees from Seoul National University, Seoul, Republic of Korea. From 2010 to 2012, he worked as a senior and principal researcher at Spatial Information Research Institute in Korea Cadastral Survey Corporation. Since 2012, he has been an Associate Professor with the school of Aerospace Engineering at Sejong University. His research interests include wide area DGNSS (WAD) correction generation algorithms, geodesy, real-time kinematics (RTK)/network RTK related algorithms, and estimation of ionospheric irregularity drift velocity using ROT variation and spaced GNSS receivers.